\documentclass[english,11pt]{amsart}
\usepackage{amsfonts,amsmath,mathrsfs,amssymb}
\usepackage{graphicx}
\usepackage[T1]{fontenc}
\usepackage{latexsym}
\usepackage{varwidth}
\usepackage{a4wide}
\usepackage{hyperref}
\usepackage{caption}
\usepackage{subcaption}

\begin{document}
\baselineskip 6mm
\date{}
\title[Coupling the NGF and MOND for cometary orbits]{Coupling the non-gravitational forces and MOdified Newton Dynamics for cometary orbits}
\author{Lucie Maquet$^{1}$}
%\address{ESA/ESAC, PO Box 78, 28691 Villanueva de la Ca\~nada, Spain}
\author{Fr\'ed\'eric Pierret$^2$}
%\address{SYRTE, Observatoire de Paris, 77 avenue Denfert-Rochereau, 75014 Paris, France}

\keywords{MOND, Non-gravitational forces, Cometary dynamics}

\begin{abstract}
In recent work (\cite{milgrom2009,bla2011,blanchet2011}), the authors showed that MOdified Newton Dynamics (MOND) have a non-negligible secular perturbation effect on planets with large semi-major axes (gaseous planets) in the Solar System. Some comets also have a very eccentric orbit with a large semi-major axis (Halley family comets) going far away from the Sun (more than 15 AU) in a low acceleration regime where they would be subject to MOND perturbation. They also approach the Sun very closely (less than 3 AU) and are affected by the sublimation of ices from their nucleus, triggering  so-called non-gravitational forces. The main goal of this paper is to investigate the effect of MOND perturbation on three comets with various orbital elements (2P/Encke, 1P/Halley and 153P/Ikeya-Zhang) and then compare it to the non-gravitational perturbations. It is motivated by the fact that when fitting an outgassing model for a comet, we have to take into account all of the small perturbing effects to avoid absorbing these effects into the non-gravitational parameters. Otherwise, we could derive a completely wrong estimation of the outgassing. For this work, we use six different forms of MOND functions and compute the secular variations of the orbital elements due to MOND and non-gravitational perturbations. We show that, for comets with large semi-major axis, the MONDian effects are not negligible compared to the non-gravitational perturbations.
\end{abstract}

\maketitle

\vskip 5mm
\begin{tiny}
\begin{enumerate}
\item {ESA/ESAC, PO Box 78, 28691 Villanueva de la Ca\~nada, Spain}

\item {SYRTE UMR CNRS 8630, Observatoire de Paris and University Paris VI, France}
\end{enumerate}
\end{tiny}

\tableofcontents

\section{Introduction}

Modified Newtonian Dynamics (MOND) has been proposed in \cite{milgrom1983_5} as an alternative to the dark matter paradigm (see \cite{sanders2002_2}). At the non-relativistic level, the best formulation of MOND is the modified Poisson equation (see \cite{bekenstein1984}),
\begin{equation}
	\label{e:MOND}
	\nabla \cdot \left[ \mu\left(\frac{g}{a_0}\right) \nabla U \right] = -4\pi G \rho\,
\end{equation}

where $\rho$ is the density of ordinary (baryonic) matter, $U$ is the gravitational potential, $\textbf{g}=\nabla U$ is the gravitational field and $g = \|\bf{g}\|$ its ordinary Euclidean norm. The modification of the Poisson equation is encoded in the MOND function $\mu(y)$ of the single argument $y\equiv g/a_0$, where $a_0=1.2\times 10^{-10}\,\mathrm{m}/\mathrm{s}^2$ denotes the MOND constant acceleration scale. The MOND function $\mu(y)$ tends to 1 for $y\gg1$ in a Newtonian strong-field regime, and tends to $y$ for $y\ll1$ in a weak gravitational field regime. According to \cite{milgrom2009}, \cite{bla2011} and \cite{blanchet2011} the most important effect of MOND in the Solar System is the External Field Effect (EFE) which produces two corrections (parametrized by two quantities $Q_2$ and $Q_4$) to the Newtonian potential which increase with the distance to the Sun. In other words, objects with a large semi-major axis are more sensitive to the effects of perturbations induced by MOND formalized by a modified Poisson equation. \\

Hence, we study comets with large semi-major axes to determine the magnitude of the effects of MOND theory. Indeed, the comets are good candidates because they not only go far from the Sun on a very eccentric orbit but also come back close to the earth to be observed accurately. When the comets approach the Sun, their gravitational orbit is affected by the sublimation of ices from their nucleus surface. The outgassing triggers non-gravitational forces that significantly modify the orbit of the comet close to the Sun (under 3 AU). These non-gravitational forces have been modeled for the first time in \cite{mar1969} and then improved in \cite{mar1973}. Other more physical approaches for the non-gravitational forces have been developed in \cite{sit1990}, \cite{sek1993}, \cite{dav2004} and \cite{maq2012}. These last models take into account outgassing from only a few areas on the nucleus which describes more accurately the observations made by space probes. \\

The model developed in \cite{mar1973} to compute the non-gravitational forces is both sufficient to study cometary orbits and more easily implemented than the more sophisticated model. This model is used to generate cometary ephemeris and gives a good estimate of the non-gravitationnal effect for cometary orbits. These non-gravitationnal forces are obtained by fitting the astrometrical data but it is important to take into account all of the small effects, such as relativistic terms, to estimate correctly the outgassing (see \cite{maqphd}). That is why the main goal of this paper is to quantify what would be the MOND perturbation on comets if this theory is validated and what is the maximum order of magnitude of this effect. \\

In \cite{hees2012_2} and \cite{hees2014_6}, the authors used the formalism developped in \cite{bla2011} and \cite{blanchet2011} to constrain the quantity $Q_2$ with the collected data of the Cassini spacecraft mission. Even though the authors claimed that the range of values of $Q_2$ are drastically restricted with that set of data, we choose to keep all the different values of $Q_2$ in order to obtain the extreme variations of the comet orbits as in \cite{bla2011} and \cite{blanchet2011}. \\

The plan of the paper is as follows : \\
 
In section \ref{pert_prob}, we present a brief reminder about the Gauss equation of the perturbed two body problem and the implementation of the non-gravitational and MOND perturbations. Section \ref{applic} shows the consequence in terms of secular variation of the orbital elements due to the non-gravitational and the MOND perturbations of three comets. We conclude in section \ref{conclu} and give some prospects.

\section{Perturbed Sun-Comet system} \label{pert_prob}

\subsection{Reminder about perturbed two-body problem}
In this Section, we recall classical definitions and results concerning the perturbed two-body problem. We refer in particular to \cite{bro1961} for more details and proofs. \\

\begin{figure}
\includegraphics[width=0.5\columnwidth]{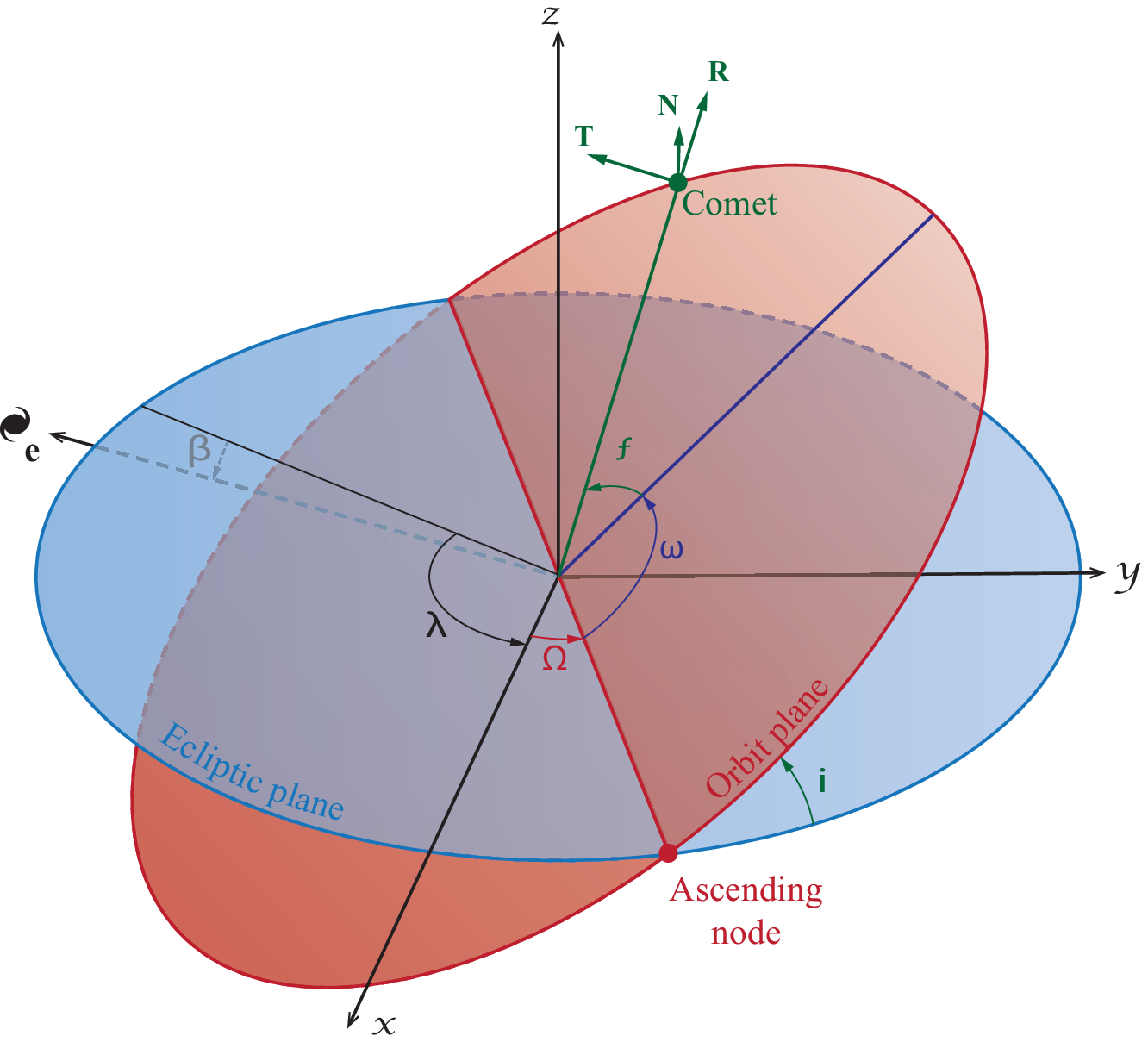}
\caption{Representation of the orbital elements as used in the text. The direction $\mathbf{e}$ is the direction of the center of the galaxy}\label{orb_fig}
\end{figure}

The unperturbed two-body problem of a comet $C$ with mass $M_C$ around the Sun $S$ with mass $M_S$ is described by six orbital elements. We adopt the most classical, which are the semi-major axis $a$, the eccentricity $e$, the orbital plane inclination $i$, the argument of the perihelion $\omega$, the ascending node longitude $\Omega$ and the mean anomaly $M$ defined by $M=n(t-\tau)$ where $n=\frac{2\pi}{P}$ is the mean motion, $P$ the orbital period of the comet and $\tau$ the time passage at the perihelion (Fig. \ref{orb_fig}). \\

Let $(S,\mathbf{x},\mathbf{y},\mathbf{z})$ be the fixed reference frame attached to the Sun, typically the fixed heliocentric frame and let $(S,\mathbf{e}_R,\mathbf{e}_T,\mathbf{e}_N)$ be the frame associated with the heliocentric motion of the comet in the orbital plane where $\mathbf{e}_R$ is the radial unit vector, $\mathbf{e}_T$ the tangential unit vector and $\mathbf{e}_N$ the normal unit vector. In $(S,\mathbf{e}_R,\mathbf{e}_T,\mathbf{e}_N)$, the position vector of the comet is written as $\mathbf{r}=r\mathbf{e}_R$. Remember that the change of basis from $(S,\mathbf{e}_R,\mathbf{e}_T,\mathbf{e}_N)$ to $(S,\mathbf{x},\mathbf{y},\mathbf{z})$ is obtained  by performing as usual three successive frame rotations with angles $\Omega$, $i$ and $f+\omega$ where $f$ is the true anomaly.\\

Let $\mathbf{a}$ be a perturbing acceleration of the comet with components $(R,T,N)$ in $(S,\mathbf{e}_R,\mathbf{e}_T,\mathbf{e}_N)$. Using the classical Gauss equations associated with the perturbed two-body problem Sun-Comet, we obtain the time variation of orbital elements as follow :

\begin{subequations}\label{gauss_eq}
\begin{eqnarray}
\frac{da}{dt} & = & \frac{2}{n\sqrt{1-e^2}}[e \sin f R + (1+e\cos f)T] \label{gauss_a}, \\ 
\frac{de}{dt} & = &\frac{\sqrt{1-e^2}}{na}[\sin f R + \left(\cos f+\frac{e+\cos f}{1+e\cos f}\right) T] \label{gauss_e},\\
\frac{di}{dt} & = & \frac{\sqrt{1-e^2} \cos (f+\omega )}{a n (1+e \cos f)}N \label{gauss_i},\\
\frac{d\Omega}{dt} & = & \frac{\sqrt{1-e^2} \sin (f+\omega )}{a n \sin i (1+ e \cos f)}N \label{gauss_Om},\\
\frac{d\omega}{dt} & = & \frac{\sqrt{(1-e^2)}}{nae}\left[-\cos f R +\left(\frac{2+e\cos f}{1+e\cos f}\right)\sin f T \right]- \cos i \frac{d\Omega}{dt}\label{gauss_om}, \\
\frac{dM}{dt} & = & n(t) - \frac{2(1-e^2)}{na^2(1+e\cos f}R -\sqrt{(1-e^2)}\left(\frac{d\omega}{dt}+\cos i \frac{d\Omega}{dt}\right) \label{gauss_M}.
\end{eqnarray}\end{subequations}

%\begin{subequations}\label{gauss_eq}
%\begin{eqnarray}
%\frac{da}{dt} & = & \frac{2}{n\sqrt{1-e^2}}[e \sin f R + (1+e\cos f)T] \label{gauss_a}, \\ 
%\frac{de}{dt} & = &\frac{\sqrt{1-e^2}}{na}[\sin f R + \left(\cos f+\frac{e+\cos f}{1+e\cos f}\right) T] \label{gauss_e},\\
%\frac{di}{dt} & = & \frac{\sqrt{1-e^2} \cos (f+\omega )}{a n (1+e \cos f)}N \label{gauss_i},\\
%\frac{d\Omega}{dt} & = & \frac{\sqrt{1-e^2} \sin (f+\omega )}{a n \sin i (1+ e \cos f)}N \label{gauss_Om},\\
%\frac{d\omega}{dt} & = & \frac{\sqrt{(1-e^2)}}{nae}\left[-\cos f R +\left(\frac{2+e\cos f}{1+e\cos f}\right)\sin f T \right] \nonumber \\
%&&- \cos i \frac{d\Omega}{dt}\label{gauss_om}, \\
%\frac{dM}{dt} & = & n(t) - \frac{2(1-e^2)}{na^2(1+e\cos f}R \nonumber \\ 
%&&-\sqrt{(1-e^2)}\left(\frac{d\omega}{dt}+\cos i \frac{d\Omega}{dt}\right) \label{gauss_M}.
%\end{eqnarray}\end{subequations}

If $\mathbf{a}$ is determined by a perturbing function $U$ as $\mathbf{a}=\nabla U$ where $\nabla$ denotes the gradient, then we have the classical relation between the perturbing force and the components $(R,T,N)$ as follow :
\begin{equation}
\label{conversion_lagrange_gauss}
R=\frac{a}{r}\frac{\partial U}{\partial a}, \ T=\frac{1}{r}\frac{\partial U}{\partial \omega} \ \text{and} \  N=\frac{1}{r\sin(f+\omega)}\frac{\partial U}{\partial i}.
\end{equation}

Using $\{c_i\}_{i=1,...,6}=\{a,e,i,\Omega,\omega,M\}$, the secular variation of the orbital elements are obtained for all $i=1,...6$ as follow :
\begin{equation}
\left< \frac{dc_i}{dt}\right>=\frac{1}{P}\int_{0}^{P} \frac{dc_i}{dt}dt=\frac{1}{2\pi}\int_{0}^{2\pi} \frac{dc_i}{dt}\frac{\left(1-e^2\right)^{3/2}}{n (1+e \cos f)^2}df.
\end{equation}

\subsection{Non-gravitational and MOND perturbations}

In \cite{mar1973}, the authors developed a semi-empirical model of the non-gravitational forces applied to comet. The illuminated surface of a spherical nucleus is assumed to be isotropically outgassing. The authors introduced the dimensionless function $g(r)$ which represents the variation in the sublimation rate as a function of the heliocentric distance of the comet. Its determination is based on the observation of the water sublimation rate curve. From the work and the model established in \cite{mar1973}, we have the non-gravitational perturbing acceleration given in $(S,\mathbf{e}_R,\mathbf{e}_T,\mathbf{e}_N)$ by its components
\begin{equation}
R_{NG}= A_1 g(r), \ T_{NG}= A_2 g(r), \ N_{NG}= A_3 g(r)
\end{equation}
where
\begin{equation}
g(r)=0.111262  \left(\frac{r}{2.808}\right)^{-2.15} \left(1+\left(\frac{r}{2.808}\right)^{5.093}\right)^{-4.6142}
\end{equation}
and $A_1, A_2, A_3$ are constants obtained by fitting the astrometrical positions of the considered comet together with the orbital elements. \\

In \cite{bla2011}, MOND was formulated to ease testing in the Solar System (for the domain with Solar distance $r\lesssim r_{0}\approx 7100 \ \mathrm{AU}$). The modification of the Newtonian gravity is given as a perturbation of the classical two-body problem. The perturbing acceleration $\mathbf{a}_{MOND}$ caused by the MOND theory in their formulation has two main perturbing parts $\mathbf{a}_{MOND,Q_2}$ and $\mathbf{a}_{MOND,Q_3}$. \cite{bla2011} show that $\mathbf{a}_{MOND,Q_3}$ is very weak for the planets. Consequently, here, we only consider $\mathbf{a}_{MOND,Q_2}$. The MOND perturbation is determined by a perturbing function given by (\cite[Eq. 40]{bla2011})
\begin{equation}
U_{MOND,Q_2} = \frac{1}{2}r^2Q_2(r)\left(\left(\mathbf{e}\cdot \mathbf{e}_R\right)^2-\frac{1}{3}\right)
\end{equation}
where $Q_2$ is a function of $r$, and $\mathbf{e}$ is the direction of the galactic center. In fact, as we consider only comets with maximum distance $r\approx100 \mathrm{AU}$, $Q_2$ is observed to be constant for $r\leq100 \ \mathrm{AU}$. Indeed, according to Fig. 4 and 5 of \cite{bla2011}, we can see that between 0 and 1000 AU, $Q_2$ varies from $3.83\times 10^{-26} \mathrm{s^{-2}}$ to $3.80\times 10^{-26} \mathrm{s^{-2}}$.\\

The values of $Q_2$ depend on the chosen MOND function $\mu$. In \cite{bla2011}, the authors deal with various MOND functions such as 
\begin{align}
\mu_n(y) &= \dfrac{y}{(1+y^n)^{1/n}}, \ \text{for any integer} \ n\ge 1, \\
\mu_{\textrm{exp}}(y) &= 1-e^{-y}, \ \mu_{\textrm{TeVeS}}(y) = \dfrac{\sqrt{1+4y} - 1}{\sqrt{1+4y} + 1}. \nonumber
\end{align}
For these MOND functions, the values of $Q_2$ are given in Table \ref{Q2values}.

\begin{table*}[h!]
\centering
\caption{Numerical values $Q_2$ for various MOND functions}. 
\label{Q2values}
\small  \begin{tabular}{lcccccc}
\hline
MOND function & $\mu_1(y)$ & $\mu_2(y)$ & $\mu_5(y)$ & $\mu_{20}(y)$ & $\mu_{\textrm{exp}}(y)$ & 
$\mu_{\textrm{TeVeS}}(y)$ \\
\hline
$Q_2$ [$\text{s}^{-2}$] & $3.8\times 10^{-26}$ & $2.2\times 10^{-26}$ & $7.4\times 10^{-27}$ & $2.1\times10^{-27}$& $3.0\times 10^{-26}$ &$4.1\times 10^{-26}$ \\
\hline
\end{tabular}
\flushleft{\textit{Note : In \cite{hees2014_6}, the authors obtained an estimated value for $Q_2=3\pm3 \times 10^{-27} \ s^{-2}$ which suggests that we consider MOND functions around the range of $\mu_2$ to $\mu_{20}$.}}
\end{table*}

In what follows, we denote the latitude and longitude of the galactic center in the heliocentric coordinate system by $\beta$ and $\lambda$ respectively, then we obtain the expression of $U_{MOND,Q2}$ as
\begin{align}
U_{MOND,Q_2} = \frac{Q_2}{6}\bigg(& 3 (x \cos \beta \cos \lambda+y \cos \beta \sin \lambda+z \sin \beta)^2 \nonumber \\
&-\left(x^2+y^2+z^2\right)\bigg),
\end{align}
where $x,y$ and $z$ are the coordinates of the comet in the frame $(S,\mathbf{x},\mathbf{y},\mathbf{z})$. In order to obtain the expression of the perturbing acceleration $\mathbf{a}_{MOND,Q_2}$ in $(S,\mathbf{e}_R,\mathbf{e}_T,\mathbf{e}_N)$, we express $x,y$ and $z$ as functions of the orbital elements using the functions in formulas \ref{conversion_lagrange_gauss} for each perturbing function. Straightforward computations lead to the expression of $R_{MOND,Q2}$, $T_{MOND,Q2}$ and $N_{MOND,Q2}$ written as 

%\begin{align}
%& R_{MOND,Q_2} = \\
%&\frac{Q_2\left(1-e^2\right) }{3(1+e \cos f)} \nonumber \\
%&\times  \bigg( 3 \big(\cos \beta (\sin \lambda (\cos i \cos \Omega \sin (f+\omega )+\sin \Omega \cos (f+\omega )) \nonumber \\
%&+\cos \lambda (\cos \Omega \cos (f+\omega )-\cos i \sin \Omega \sin (f+\omega ))) \nonumber \\
%&+\sin \beta \sin i \sin (f+\omega )\big)^2-1\bigg), \nonumber\\
%& T_{MOND,Q_2}= \\
%&-\frac{Q_2 a(1-e^2)}{1+e\cos f} \nonumber\\
%&\times\bigg(\cos (\beta ) \sin (f+\omega ) \cos (\lambda -\Omega )\nonumber \\
%&\quad \quad -\cos (f+\omega ) (\cos (\beta ) \cos (i) \sin (\lambda -\Omega )+\sin (\beta ) \sin (i))\bigg) \nonumber \\
%& \times \bigg(\cos (\beta ) (\cos (i) \sin (f+\omega ) \sin (\lambda -\Omega )\nonumber \\
%&\quad \quad +\cos (f+\omega ) \cos (\lambda -\Omega ))+\sin (\beta ) \sin (i) \sin (f+\omega )\bigg),\nonumber\\
%& N_{MOND,Q_2}=\\
%&\frac{Q_2a(1-e^2)}{1+e\cos f}(\sin (\beta ) \cos (i)-\cos (\beta ) \sin (i) \sin (\lambda -\Omega )) \nonumber \\
%&\times\bigg(\cos (\beta ) (\cos (i) \sin (f+\omega ) \sin (\lambda -\Omega )\nonumber \\
%&\quad \quad+\cos (f+\omega ) \cos (\lambda -\Omega ))+\sin (\beta ) \sin (i) \sin (f+\omega )\bigg).\nonumber
%\end{align}

\begin{align}
R_{MOND,Q_2}&=\frac{Q_2\left(1-e^2\right) }{3(1+e \cos f)} \nonumber \\
&\times  \bigg( 3 \big(\cos \beta (\sin \lambda (\cos i \cos \Omega \sin (f+\omega )+\sin \Omega \cos (f+\omega )) \nonumber \\
&+\cos \lambda (\cos \Omega \cos (f+\omega )-\cos i \sin \Omega \sin (f+\omega ))) \nonumber \\
&+\sin \beta \sin i \sin (f+\omega )\big)^2-1\bigg),\\
T_{MOND,Q_2}=&-\frac{Q_2 a(1-e^2)}{1+e\cos f} \nonumber\\
&\times\bigg(\cos (\beta ) \sin (f+\omega ) \cos (\lambda -\Omega )\nonumber \\
&\quad \quad -\cos (f+\omega ) (\cos (\beta ) \cos (i) \sin (\lambda -\Omega )+\sin (\beta ) \sin (i))\bigg) \nonumber \\
& \times \bigg(\cos (\beta ) (\cos (i) \sin (f+\omega ) \sin (\lambda -\Omega )\nonumber \\
&\quad \quad +\cos (f+\omega ) \cos (\lambda -\Omega ))+\sin (\beta ) \sin (i) \sin (f+\omega )\bigg),\\
N_{MOND,Q_2}=&\frac{Q_2a(1-e^2)}{1+e\cos f}(\sin (\beta ) \cos (i)-\cos (\beta ) \sin (i) \sin (\lambda -\Omega )) \nonumber \\
&\times\bigg(\cos (\beta ) (\cos (i) \sin (f+\omega ) \sin (\lambda -\Omega )\nonumber \\
&\quad \quad+\cos (f+\omega ) \cos (\lambda -\Omega ))+\sin (\beta ) \sin (i) \sin (f+\omega )\bigg).
\end{align}

We now have all the expressions of the perturbing force so we can deduce the secular variations of the orbital elements of the comet which are given as follows :
\begin{equation}
\left< \frac{dc_i}{dt}\right>=\left< \frac{dc_i}{dt}\right>_{NG}+\left< \frac{dc_i}{dt}\right>_{MOND,Q_2}
\end{equation}
for all $i=2,...,6$. For the secular variation of the semi-major axis we only have
\begin{equation}
\left< \frac{da}{dt}\right>=\left< \frac{da}{dt}\right>_{NG}
\end{equation}
because the secular variation caused by MOND on the semi-major axis is zero.

\section{Application to three comets} \label{applic}
We now compute the effects of the non-gravitational and MOND perturbations on three comets. We computed analytically the MONDian part and numerically the non-gravitational part due to the expression of the equations.\\

The three comets were choosen because of their orbital parameters. As the MONDian effects are bigger for objects far from the sun, we choose 1P/Halley and 153P/Ikeya-Zhang which have amongst the largest semi-major axes known for periodic comets. They are also relatively well know (8154 and 1954 astrometrical observations respectively). Conversely, we choose 2P/Encke for its small semi-major axis to be able to make a comparison of the two types of comet. \\

In Table \ref{tab:orb_el_JPL}, we present truncated values of the orbital elements and non-gravitational parameters. These orbital elements are given by the database of JPL Small-Bodies Browser\footnote{\url{http://ssd.jpl.nasa.gov/sbdb.cgi}}. For the computations, we used non-truncated values of these elements. We refer to the JPL Small-Bodies Browser website for the values and their uncertainties. It can be noted that, for most of the comets, the non-gravitational parameter $A_3$ is considered as zero. Indeed, the non-gravitational perturbation in this direction is very weak and cannot be solved by the fit (see \cite{maqphd} and \cite{mar1973}).\\

Using the values of the latitude $\beta$ and longitude $\lambda$ of the galactic center in the fixed heliocentric reference frame which are $\beta=-5.5^\circ$ and $\lambda=-93.2^\circ$ (see for example \cite{all2000}), the results of the computation are given in Tables \ref{tab:res_encke}, \ref{tab:res_halley} and \ref{tab:res_ikeya}. The secular variation of the angles induced by MOND are in the range of a few \textit{milli}-arc-seconds per century. We can note that for 2P/Encke (short orbit), the MONDian effects are very small and may be negligible compared to the non-gravitational perturbation. Conversely, for 1P/Halley and 153P/Ikeya-Zhang (large orbit), the MONDian effects are much bigger and must be included together with the non-gravitational perturbations. As noticed in \cite{bla2011}, the effects of the MOND perturbations decreases by a factor $\approx 10$ for the MOND function $\mu_n$ between $n=2$ and $n=20$. But the effects are in the same range for $\mu_1$, $\mu_{exp}$ and $\mu_{TeVeS}$. \\

According to Table \ref{tab:res_encke}, \ref{tab:res_halley} and \ref{tab:res_ikeya}, the cometary orbits are precessing under non-gravitational perturbations but also because of the modified dynamics. As the non gravitational parameter $A_3$ is zero, there is no secular variation of the inclination of the orbit or of the longitude of the ascending node due to the outgassing from the comet. The variation of these elements is purely due to the modified dynamics. We also computed the effects of MOND on the eccentricity but it was not significant ($\left<\frac{de}{dt}\right>$ takes a value around $10^{-10}$-$10^{-11}$ $\mathrm{cy}^{-1}$).\\

For the well-known comet 1P/Halley (seen by the ESA spacecraft Giotto in 1986), the MONDian effects are in the same range as the precision of the orbital element determination ($\sigma_i=24.4$ mas, $\sigma_{\omega}=42.2$ mas and $\sigma_{\Omega}=32.6$ mas, see JPL Small Bodies browser website for more details). As the dynamical models of comets are continually improving, it will soon be possible to detect and quantify these effects in cometary orbits. \\

We also computed the perturbation induced by $\textbf{a}_{MOND,Q3}$. To give an idea of the effect, it is about $10^{-11}$ to $10^{-12}$ mas/cy for 153P/Ikeya-Zhang and much less for the two other comets. With the precision of current observations, it is completely negligible.

\begin{table*}[htbf!]
\centering
\caption{Truncated values of the orbital elements and non-gravitational parameters of the comets from the JPL small bodies browser. For the computations, we used non-truncated values available on the JPL website.}
\label{tab:orb_el_JPL}
\begin{tabular}{ cccc }
\hline
& 2P/Encke & 1P/Halley & 153P/Ikeya-Zang \\
\hline
P [yr] & 3.30 & 75.31 & 366.51 \\
a [AU] & 2.215 & 17.834 & 51.214 \\
e & 0.848 & 0.967 & 0.990 \\
i [deg]  & 11.8 & 162.3 & 28.1 \\
$\omega$ [deg] & 186.5 & 111.3 & 34.7 \\
$\Omega$ [deg] & 334.6 & 58.4 & 93.4 \\
n [deg.$\mathrm{day}^{-1}$] & 0.299 & 0.013 & 0.003 \\
q [AU] & 0.336 & 0.586 & 0.507 \\
$\mathrm{A_1}$ [AU.$\mathrm{day}^{-2}$] & $1.58\times10^{-10}$ & $2.70\times10^{-10}$  & $3.33\times10^{-9}$ \\
$\mathrm{A_2}$ [AU.$\mathrm{day}^{-2}$] & $-5.05\times10^{-11}$ & $1.56\times10^{-10}$ & $-3.51\times10^{-10}$ \\
$\mathrm{A_3}$ [AU.$\mathrm{day}^{-2}$] & $0.0$ & $0.0$ & $0.0$\\
\hline
\end{tabular}
\end{table*}

\begin{table*}[htbf!]
\begin{center}
\caption{Results for the secular variations of 2P/Encke due to the MOND and non-gravitational perturbations. $\left<\frac{da}{dt}\right>$ is given in astronomical ($\text{AU}/\text{cy}$), $\left<\frac{de}{dt}\right>$ is given in $\text{cy}^{-1}$ and all results for the angles are given in \textit{milli}-arc-seconds per century ($\text{mas}/\text{cy}$).}
\label{tab:res_encke}
\begin{tabular}{cccccc}
\hline
MOND function & $\left<\frac{da}{dt}\right>$ & $\left<\frac{de}{dt}\right>$  & $\left<\frac{di}{dt}\right>$ & $\left<\frac{d\Omega}{dt}\right>$ & $\left<\frac{d\omega}{dt}\right>$ \\
\hline
$\mu_1(y)$ & - & - & 0.125 & -0.062 & -0.23 \\
$\mu_2(y)$ & - & - & 0.072 & -0.036 & -0.13 \\
$\mu_5(y)$ & - & - & 0.024 & -0.012 & -0.04 \\
$\mu_{20}(y)$ & - & - & 0.007 & -0.003 & -0.01 \\
$\mu_{\textrm{exp}}(y)$ & - & - & 0.099 & -0.049 & -0.18 \\
$\mu_{\textrm{TeVeS}}(y)$ & - & - & 0.135 & -0.067 & -0.25 \\
\hline
\hline
Non-gravitational & $-6.76 \times 10^{-5}$ & $-3.87 \times 10^{-6}$ & - & - & -3477.51 \\
\hline
\end{tabular}
\end{center}
\end{table*}

\begin{table*}[htbf!]
\begin{center}
\caption{Results for the secular variations of 1P/Halley due to the MOND and non-gravitational perturbations. $\left<\frac{da}{dt}\right>$ is given in astronomical ($\text{AU}/\text{cy}$), $\left<\frac{de}{dt}\right>$ is given in $\text{cy}^{-1}$ and all results for the angles are given in \textit{milli}-arc-seconds per century ($\text{mas}/\text{cy}$).}
\label{tab:res_halley}
\begin{tabular}{cccccc}
\hline
MOND function & $\left<\frac{da}{dt}\right>$ & $\left<\frac{de}{dt}\right>$  & $\left<\frac{di}{dt}\right>$ & $\left<\frac{d\Omega}{dt}\right>$ & $\left<\frac{d\omega}{dt}\right>$ \\
\hline
$\mu_1(y)$ & - & - & 3.06 & -33.95 & -19.07 \\
$\mu_2(y)$ & - & - & 1.77 & -19.66 & -11.04 \\
$\mu_5(y)$ & - & - & 0.60 & -6.61 & -3.71  \\
$\mu_{20}(y)$ & - & - & 0.17 & -1.88 & -1.05  \\
$\mu_{\textrm{exp}}(y)$ & - & - & 2.41 & -26.80 & -15.05 \\
$\mu_{\textrm{TeVeS}}(y)$ & - & - &  3.30 & -36.63 & -20.572 \\
\hline
\hline
Non-gravitational & $2.62 \times 10^{-3}$ & $4.33 \times 10^{-6}$ & - & - & -508.72 \\
\hline
\end{tabular}
\end{center}
\end{table*}

\begin{table*}[htbf!]
\begin{center}
\caption{Results for the secular variations of 153P/Ikeya-Zang due to the MOND and non-gravitational perturbations. $\left<\frac{da}{dt}\right>$ is given in astronomical ($\text{AU}/\text{cy}$), $\left<\frac{de}{dt}\right>$ is given in $\text{cy}^{-1}$ and all results for the angles are given in \textit{milli}-arc-seconds per century ($\text{mas}/\text{cy}$).}
\label{tab:res_ikeya}
\begin{tabular}{cccccc}
\hline
MOND function & $\left<\frac{da}{dt}\right>$ & $\left<\frac{de}{dt}\right>$  & $\left<\frac{di}{dt}\right>$ & $\left<\frac{d\Omega}{dt}\right>$ & $\left<\frac{d\omega}{dt}\right>$ \\
\hline
$\mu_1(y)$ & - & - & -10.57 & -45.30 & -43.41 \\
$\mu_2(y)$ & - & - & -6.12 & -26.22 & -25.13 \\
$\mu_5(y)$ & - & - & -2.06 & -8.82 & -8.45 \\
$\mu_{20}(y)$ & - & - & -0.58 & -2.50 & -2.40 \\
$\mu_{\textrm{exp}}(y)$ & - & - & -8.34 & -35.76 & -34.27 \\
$\mu_{\textrm{TeVeS}}(y)$ & - & - & -11.40 & -48.87 & -46.83 \\
\hline
\hline
Non-gravitational & 0.17 & $3.23\times 10^{-5}$ & - & - & -5168.5 \\
\hline
\end{tabular}
\end{center}
\end{table*}

\section{Conclusion}\label{conclu}
This work shows that the effects of MOND theory are not negligible compared with other small perturbations like non-gravitational perturbations on the cometary orbits. In agreement with the study of \cite{bla2011}, the MONDian effects are stronger for large orbits than for short orbits.\\

If the MOND theory is validated, it is really important to take into account its effects on the secular variation of cometary orbital elements, especially in the case of Halley family comets with large orbits. Indeed, as for the relativistic terms, the MOND perturbation would not be absorbed in the fitting of the constants $A_1$, $A_2$ and $A_3$ representing the outgassing. \\

Some long-term studies of cometary orbits have already begun, for example to constrain the Oort cloud density through incoming new comets from this cloud (for example \cite{fou2011}). Generally, this kind of study takes into account the galactic tide, the star encounters and the non-gravitational effects of the new comets introduced into the inner Solar System. It would be interesting to include MONDian perturbations to these studies on the injection of new comets into the Solar System and hence improve the Oort object density.\\

Finally, thanks to space missions like Rosetta (see \cite{sch2012}), our understanding of cometary physics will be improved in the very near future and consequently the cometary dynamical model will be more accurate. In this way, the models of outgassing such as \cite{mar1973} or \cite{maq2012} including MOND perturbation will be tested through ephemeris computation in order to validate (or otherwise) such a perturbation. This work will the subject of a future paper.

\section*{Acknowledgements}
It is a great pleasure to thank Luc Blanchet for a careful reading and interest in the manuscript and very useful comments and advice. We also thank Tim Rawle for his careful proof reading.\\

Lucie Maquet is supported by a Europeen Space Agency (ESA) research fellowship at the Europeen Space Astronomy Center (ESAC), in Madrid, Spain.

\bibliographystyle{plain}
\bibliography{MP_MOND_NG}

\end{document}